\documentclass[aps,prl,twocolumn,showpacs,superscriptaddress]{revtex4}
\usepackage{graphicx}
\usepackage{amssymb}

\begin{document}

\title{Magneto-structural transitions in a frustrated magnet at high fields }

\author{V. Tsurkan}

\affiliation{Experimental Physics 5, Center for Electronic Correlations and Magnetism, Institute of Physics, University of Augsburg, D 86159, Augsburg, Germany}

\affiliation{Institute of Applied Physics, Academy of Sciences of Moldova, MD 2028, Chisinau, R. Moldova}

\author{S. Zherlitsyn}

\affiliation{Hochfeld-Magnetlabor Dresden (HLD), Helmholtz-Zentrum Dresden-Rossendorf, D-01314 Dresden, Germany}

\author{V. Felea}

\affiliation{Institute of Applied Physics, Academy of Sciences of Moldova, MD 2028, Chisinau, R. Moldova}

\affiliation{Institute for Condensed Matter Physics, TU Braunschweig, D-38106 Braunschweig, Germany}

\author{S. Yasin}
\author{Yu. Skourski}

\affiliation{Hochfeld-Magnetlabor Dresden (HLD), Helmholtz-Zentrum Dresden-Rossendorf, D-01314 Dresden, Germany}

\author{J. Deisenhofer}
\author{H.-A. Krug von Nidda}

\affiliation{Experimental Physics 5, Center for Electronic Correlations and Magnetism, Institute of Physics, University of Augsburg, D 86159, Augsburg, Germany}

\author{P. Lemmens}

\affiliation{Institute for Condensed Matter Physics, TU Braunschweig, D-38106 Braunschweig, Germany}

\author{J. Wosnitza}

\affiliation{Hochfeld-Magnetlabor Dresden (HLD), Helmholtz-Zentrum Dresden-Rossendorf, D-01314 Dresden, Germany}

\author{A. Loidl}

\affiliation{Experimental Physics 5, Center for Electronic Correlations and Magnetism, Institute of Physics, University of Augsburg, D 86159, Augsburg, Germany}

\date{24.05.2011 [Received: date / Revised version: date ]}

\begin{abstract}

Ultrasound and magnetization studies of bond-frustrated ZnCr$_2$S$_4$ spinel are performed in static magnetic fields up to 18~T and in pulsed fields up to 62~T. At temperatures below the antiferromagnetic transition at $T_{N1}\approx$ 14~K the sound velocity as function of magnetic field reveals a sequence of steps followed by plateaus indicating a succession of crystallographic structures with constant stiffness. At the same time, the magnetization evolves continuously with field up to full magnetic polarization without any plateaus in contrast to geometrically frustrated chromium oxide spinels. The observed high-field magneto-structural states are discussed within a \emph{H-T}~phase diagram taking into account the field and temperature evolution of three coexisting spin structures and subsequent lattice transformations induced by magnetic field.

\end{abstract}

\pacs{43.35.+d, 62.65.+k, 72.55.+s, 75.50.Ee }

\maketitle

Frustrated magnets with spinel structure $AB_2X_4$ manifest an intriguing behavior and unusual ground states, such as composite spins \cite{Lee02}, spin dimerization \cite{Radaelli02, Schmidt04}, heavy-fermion properties \cite{Kondo97,Krimmel99}, spin-orbital liquid \cite{Fritsch04}, and orbital glass \cite{Fichtl05,Tsurkan05} which originate from magnetic frustration but also from the intimate interplay of spin, charge and orbital degrees of freedom and their coupling to the lattice. In the magnetic \emph{B}-site Cr spinels with strong spin-phonon coupling, a novel type of structural transformation has been identified experimentally, the so called spin Jahn-Teller effect \cite{Lee00,Sushkov05,Hemberger06,Hemberger07}. In an octahedral crystal field the \emph{t}$_{2g}$ levels of the Cr$^{3+}$ ions are half filled and the spin-orbit coupling is negligible. Therefore, the conventional Jahn-Teller scenario related to magnetic ions with an orbitally degenerate state is not applicable here, and the structural deformation is believed to be driven purely by spin ordering. The ground-state properties of frustrated magnets are characterized by a large degeneracy and are highly susceptible to external perturbations. An external magnetic field can change the balance between the competing interactions, and unusual phenomena, such as magnetization plateaus at half or fractional saturation are observed \cite{Ueda05,Matsuda07}.

In geometrically frustrated $A$Cr$_2X_4$ oxide ($X$=O) spinels the Cr ions forming a pyrochlore lattice of corner-sharing tetrahedra are strongly coupled by direct antiferromagnetic (AFM) interactions of the order of 100 - 400 K. Emerging phenomena in geometrically frustrated oxides are dominated by local "tetrahedron" physics \cite{Yamashita00,Tchernyshyov02}. In sulphide ($X$=S) and selenide ($X$=Se) spinels the direct AFM exchange is reduced due to the increasing distance between the magnetic ions and at the same time 90$^o$ ferromagnetic (FM) exchange becomes important. Where FM and AFM exchanges are of comparable strength, the ground state again is strongly frustrated, a situation which has been named bond frustration \cite{Rudolf07}.

In ZnCr$_2$S$_4$, subject of the present study, competing FM and AFM interactions indeed are of equal strength resulting in a Curie-Weiss temperature close to zero \cite{Hemberger06}. Neutron-diffraction measurements \cite{Hamedoun86a, Hamedoun86b,Yokaichiya09} established two subsequent magnetic transitions in ZnCr$_2$S$_4$: the first one to an incommensurate helical AFM order at $T_{N1}\approx$ 14~K,
and the second one, to coexisting commensurate spin order at $T_{N2}\approx$ 7~K. The helical state is characterized
by a spin spiral with a propagation vector \emph{$k_{1}$} $\approx(0, 0, 0.787)$ along the crystallographic \emph{c} direction and with
the spins rotating in the \emph{a-b} plane. This is also the ground-state structure of the bond-frustrated AFM ZnCr$_2$Se$_4$ with increased FM exchange as compared to ZnCr$_2$S$_4$ \cite{Hemberger07}. At temperatures below $T_{N2}$ the spiral phase coexists with two additional collinear spin structures with propagation vectors \emph{$k_{2}$} $\approx (0.5, 0.5, 0)$ and
\emph{$k_{3}$} $\approx (1.0, 0.5, 0)$ \cite{Yokaichiya09}. These collinear ordering wave vectors resemble those of geometrically frustrated AFM ZnCr$_2$O$_4$ which exhibits composite spin structures of weakly interacting self-organized spin clusters \cite{Lee02,Ji09}. The magnetic ground state of ZnCr$_2$S$_4$ can be regarded as a combination of spin orders known from geometrically frustrated ZnCr$_2$O$_4$ and bond frustrated ZnCr$_2$Se$_4$. An external field favors the parallel spin alignment and one can expect that the system passes through a sequence of exotic states when frustration is released via strong magneto-elastic coupling.

Several bulk properties of ZnCr$_2$S$_4$, such as specific heat and thermal expansion exhibit significant anomalies at the magnetic transitions. Also, a pronounced splitting of the phonon modes in the IR reflectivity spectra below the magnetic transitions was found clearly indicative for broken symmetry \cite{Hemberger06}. These anomalies suggest structural transformations due to a strong spin-phonon coupling \cite{Rudolf07}. Recent high-resolution synchrotron x-ray powder diffraction measurements  indeed revealed two subsequent structural transformations in ZnCr$_2$S$_4$ from a cubic \emph{Fd$\bar{3}$m} to a tetragonally distorted intermediate phase (space group \emph{$I4_1$/amd}) below $T_{N1}$ and a further transition into a low-temperature orthorhombic phase (space group \emph{Imma}) below $T_{N2}$ \cite{Yokaichiya09}.

In this Letter, we report on ultrasound and magnetization studies on ZnCr$_2$S$_4$ single crystals performed in static (up to 18~T) and pulsed magnetic fields (up to 62~T). We explore the unique case of almost fully compensated AFM and FM exchange interactions in the spinel ZnCr$_2$S$_4$ with strong spin-lattice coupling and several coexisting magnetic structures with different spin arrangement. We expected that strong spin-phonon coupling leads to significant fingerprints in the temperature and magnetic field dependence of sound waves. Ultrasound techniques are known to be highly sensitive probes for magneto-elastic interactions \cite{Lüthi05}. It has been demonstrated by L\"{u}thi \emph{et al.} that the magnetic phase transition in ZnCr$_2$O$_4$, recognized now as spin Jahn-Teller effect, is accompanied by pronounced anomalies in the temperature dependence of sound waves \cite{Kino71}.

ZnCr$_2$S$_4$ single crystals were grown by chemical transport reactions.
The magnetic susceptibility reveals a sharp peak at $T_{N1} =$ 13.8~K in good agreement with published data \cite{Hemberger06}. The
measurements of the velocity and attenuation of longitudinal waves with wave vector \textbf{k} and polarization \textbf{u} parallel to the $\langle001\rangle$ axis (corresponding to \emph{c}$_{11}$ acoustic mode for a cubic crystal) were performed for temperatures between 1.5 and 300~K and in static magnetic fields utilizing an experimental setup similar to that in \cite{Wolf01} with a phase-sensitive detection technique based on a pulse-echo method. The measurements in pulsed fields with a rise time of 35~ms and a pulse duration of 150~ms were done in the range 1.5 -- 20~K.

In Fig.~1 the relative changes of the sound velocity and attenuation for different static magnetic fields are presented as function of temperature. In zero field, the sound velocity exhibits significant softening on decreasing temperature below 60 K and a well-defined anomaly at $T_{N1} =$ 13.8~K, in agreement with susceptibility and thermal-expansion data \cite{Hemberger06}. The sound attenuation raises sharply approaching $T_{N1}$ and exhibits a well-defined peak. With increasing magnetic field the magnitude of the anomalies in the sound velocity and attenuation at $T_{N1}$ are reduced and shifted to lower temperatures. We recall that at $T_{N1}$ the sample transforms from the cubic paramagnetic state into the tetragonal helimagnetic phase \cite{Yokaichiya09}. The softening of the sound velocity probably results from strong spin fluctuations in the cooperative paramagnetic state where the thermal energy is lower than the leading frustrated magnetic exchange. In contrast, at $T_{N2}$ = 7.3~K where ZnCr$_2$S$_4$ transforms into the orthorhombic structure, only a smooth change of the sound velocity is found, and a broad peak in the attenuation appears. According to \cite{Hamedoun86a} the collinear structure evolves already at 12~K which could explain the observation of two peaks in the attenuation below $T_{N1}$ and of a weak anomaly in the sound velocity at $T_{N2}$. The anomaly in the attenuation at 11~K shifts to lower temperatures with increasing fields. Its magnitude first increases with field up to 5~T but then decreases at higher fields.

\begin{figure}[t]
\centering
\includegraphics[angle=0,width=0.35\textwidth]{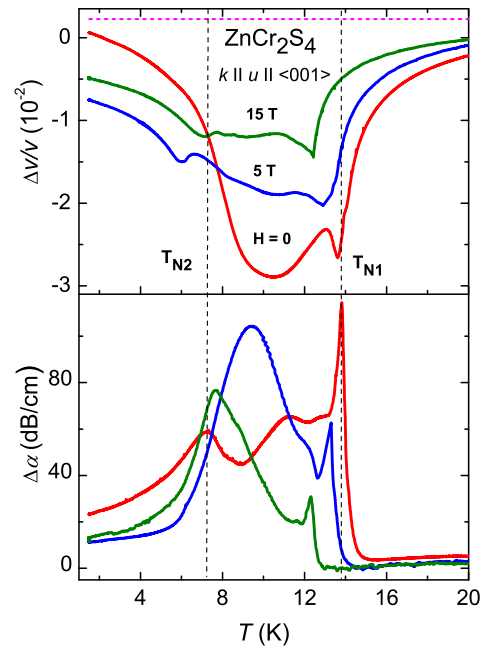}
 \caption{(color online) Temperature dependencies of the relative change of the sound velocity $\Delta \emph{v/v$_0$}$ (upper panel) and sound attenuation $\Delta \alpha$ (lower panel) for ZnCr$_2$S$_4$ measured in different static magnetic fields. The vertical dashed lines mark the magnetic phase transitions $T_{N1}$ and $T_{N2}$ in zero field. The horizontal dashed line shows the extrapolated undisturbed sound velocity estimated from a fit to the data above 60~K (see text for details).}
\end{figure}

\begin{figure*}[htb]
\includegraphics[angle=0,width=0.95 \textwidth]{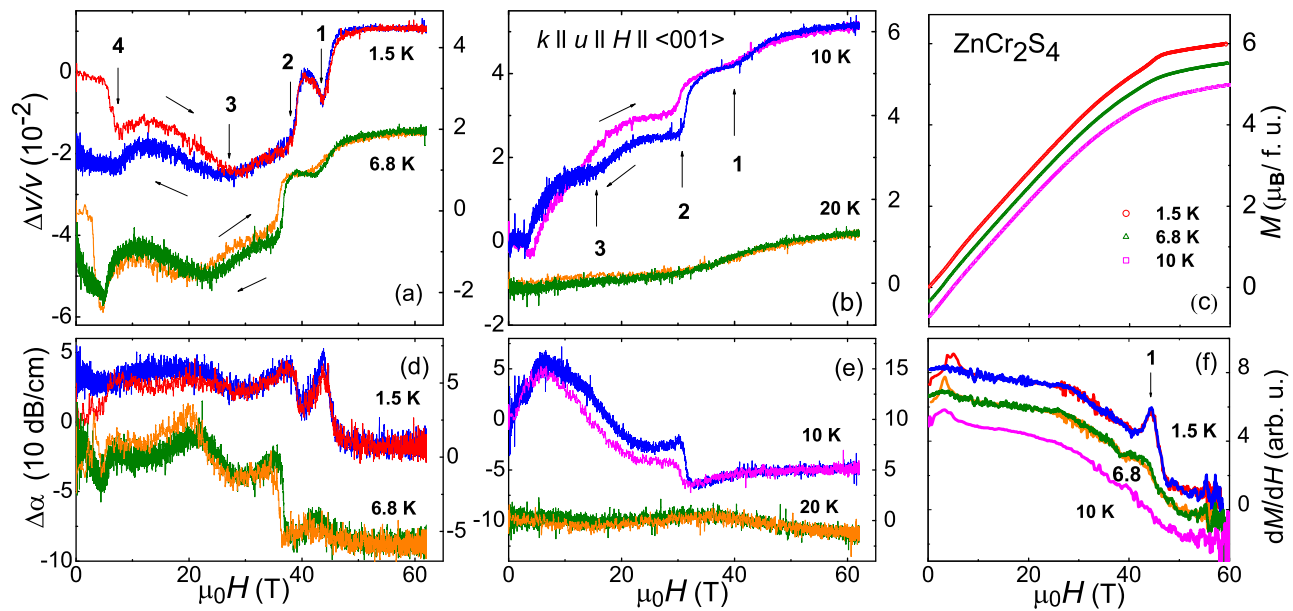}
 \caption{(color online) Relative change of the sound velocity,
    $\Delta$\emph{v/v}$_0$ [(a) and (b)] and attenuation,
  $\Delta \alpha$ [(d) and (e)] in  ZnCr$_2$S$_4$ vs. magnetic field at 1.5~K, 10~K (left scale)  and 6.8~K, 20~K (right scale).
   The vertical arrows mark the magneto-structural anomalies labeled from 1 to 4.  (c)
    Magnetization curves with field aligned along the $\langle 001\rangle$ axis and (f) derivatives of
     the magnetization d\emph{M}/d\emph{H} for 1.5, 6.8, and 10~K. For clarity the curves are shifted  along the vertical axis.
     Data for field sweep up and down are shown.}
\end{figure*}

Figure~2 documents the main results of the pulsed field studies presenting the relative change of the sound velocity,
$\Delta \emph{v/v$_0$}$, attenuation, $\Delta \alpha$, and magnetization,\emph{ M}, as function of applied field for different temperatures.
We notice rather significant changes of the sound velocity with field of the order of 3-5\% which prove the strong magneto-elastic coupling in ZnCr$_2$S$_4$. At
1.5~K, the sound velocity shows a non-monotonous behavior with field; $\Delta \emph{v/v$_0$}$ first decreases and beyond 30~T increases again with increasing fields.
Both, the sound velocity and the attenuation exhibit four prominent anomalies at 7, 27, 38, and 44~T suggesting changes in the spin state and structural phase transitions. The anomalies (labeled  respectively from 4 to 1) are visible as minima or clear changes of slope in the sound velocity, and as maxima in the attenuation. The anomalies 1-3 (at 1.5~K) in the sound velocity at fields above 27~T are free of hysteresis, whereas anomaly 4 shows a marked hysteresis on increasing and decreasing fields. Such hysteretic behavior indicates first-order transformations induced by the magnetic field. With increasing temperature all anomalies in the sound velocity shift to lower fields. At 10~K, the sound velocity develops into distinct plateaus with step-like features at 16.5, 30, and 40~T. This signals abrupt changes of the lattice stiffness, probably induced by structural phase transitions, which are followed by plateaus with a given structure and, therefore, constant stiffness [Fig.~2(b)]. At the same time, between 7.5 and 40~T the changes of the magnetization \emph{M} with field are rather gradual [Fig.~2(c)] and occur with two different slopes below and above 27~T [Fig. 2(f)]. This indicates that the structural changes are not accompanied by significant changes in the spin structure. In Fig.~2(f) the field derivatives of the magnetization d\emph{M}/d\emph{H} for different temperatures are shown. The sharp maximum in d\emph{M}/d\emph{H} at 44~T (at 1.5~K) correlates well with the anomaly 1 in $\Delta \emph{v/v$_0$}$ and $\Delta \alpha$. Finally, at 47~T the full saturated polarization is achieved with the net ordered moment close to $6~\mu_{B}$. A well-defined change of slope in \emph{M} appears just before reaching full polarization. With increasing temperature, the maximum in d\emph{M}/d\emph{H} is shifted to lower fields and broadens considerably.

The magnetic-field dependence of the sound velocity and attenuation below 15~T is dominated by dynamic effects as revealed by measurements in static fields.
A comparison of $\Delta \emph{v/v$_0$}$ in pulsed and static fields shows a general agreement. However, below 7~T the variations of $\Delta \emph{v/v$_0$}$ in static fields are much smaller than in pulsed fields which can be attributed to the relaxation dynamics of domains reorientations that should be comparable to the high sweep rate in pulsed fields.

The observed temperature and field evolution of the anomalies in the sound characteristics detected both in static and pulsed fields is summarized in a tentative  phase diagram in Fig.~3. We interpret it within the scenario which considers the interplay of different magnetic phases with structural transformations. In zero field at \emph{T} = 1.5~K (phase V), two commensurate collinear (\emph{$k_{2}$}+\emph{$k_{3}$}) spin structures coexist with an incommensurate helical (\emph{$k_{1}$}) spin structure and the crystal structure is orthorhombic \cite{Yokaichiya09}. At fields above 7.5~T (phase IV), the commensurate structure (\emph{$k_{3}$}) becomes suppressed but the second commensurate structure (\emph{$k_{2}$}) survives and coexists with the helical spin structure (\emph{$k_{1}$}). At the same time the suppression of the splitting of the lowest phonon by this field strength observed in the IR experiments \cite{Hemberger06,Rudolf07} indicates the change of the lattice symmetry, from orthorhombic to tetragonal which identifies the origin of the anomaly 4 in the ultrasound data.

The transition from the phase IV into the phase III can be traced by the anomaly in the attenuation (Fig.~1, static data) and by the anomaly 3 in the pulsed-field scans in Fig.~2. Above 7~K in the static fields this transition is clearly defined, whereas in the pulsed fields the boundary between phases IV and III is much broader probably due to relaxation effects. But we cannot exclude an intermediate phase in between phases III and IV for temperatures 7~K~$<$~\emph{T}~$<$~12~K. The phase III corresponds to a tetragonal phase with the Cr spins forming a spiral, such as in ZnCr$_2$Se$_4$.

\begin{figure}[htb]
\includegraphics[angle=0,width=0.39\textwidth]{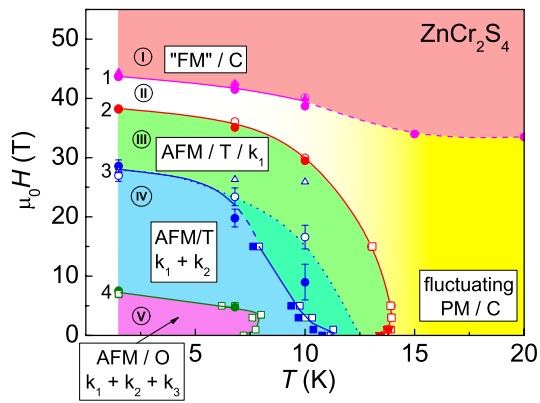}
 \caption{(color online) \emph{H-T} phase diagram of ZnCr$_2$S$_4$. Circles correspond to ultrasound data for pulsed fields, squares for static fields; up triangles to magnetization data for pulsed fields, down triangles for static fields. Open symbols for up sweeps in field (temperature), closed symbols for down sweeps. Anomalies 1-4 and different magneto-structural phases I-V are described in the text.   }
\end{figure}

The best defined features in the sound velocity and in the attenuation both in temperature and field dependencies are reflected by anomaly 2. Its step-like shape is only slightly affected by temperature indicating a well-defined phase transformation. Since the magnetization shows only a gradual change in this temperature and field range, and taking into account neutron diffraction and x-ray data \cite{Hamedoun86a,Hamedoun86b,Yokaichiya09} we associate the anomaly 2 with the second structural transition from the tetragonal phase III to the cubic phase II. It is also important to note that the magnitude of the total change of $\Delta \emph{v/v$_0$}$ by magnetic field is close to that expected for the full recovery of the cubic state estimated from a fit to the experimental data in the true paramagnetic state (above 60 K) shown by the dashed line in Fig.~1, using an anharmonic approximation according to Ref. 25. We suppose that at zero temperature in fields close to 40~T the AFM spin spiral arrangement becomes fully suppressed and the system enters into a strongly polarized cubic paramagnetic state because no additional phase boundary to the paramagnetic phase above $T_{N1}$ is evidenced in the ultrasound data. The anomaly 1 in the ultrasound and magnetization data  probably has a purely magnetic origin. Finally, phase I corresponds to a state with full polarization induced by magnetic field.

In conclusion, ultrasound and magnetization studies in magnetic fields up to 62~T of bond-frustrated ZnCr$_2$S$_4$ with strong magneto-elastic coupling revealed a sequence of magneto-structural states. We evidenced novel effects, namely, plateaus in the sound velocity on the way towards the recovery of the lattice symmetry in the polarized state which are ascribed to different crystallographic phases with constant stiffness. In contrast to geometrically frustrated antiferromagnets $A$Cr$_2$O$_4$ ($A$= Zn, Cd, Hg) which reveal magnetization plateaus accompanied by lattice distortions \cite{Ueda05,Matsuda07,Zherlitsyn10,Kojima10,Mitamura07}, the magnetization of bond-frustrated ZnCr$_2$S$_4$ evolves continuously without any anomalies up to full polarization. Our study provides a new insight into the physics of bond-frustrated spinels which is clearly distinct from that of the geometrically frustrated oxides. The origin of the observed intriguing effects, in particular, of the anomalies 1-3, as well as of the coupling mechanism of different magnetic structures to lattice strain which generates crystal structures of different symmetry is yet to be clarified and demands for further experimental and theoretical studies.

\begin{acknowledgements}

This research has been supported by the DFG via TRR 80 (Augsburg - Munich) and FOR 960 (Quantum phase transitions), LE 967/6-1 (Braunschweig) and by EuroMagNET II under the contract 228043.

\end{acknowledgements}


\begin{thebibliography}{99}

\bibitem{Lee02} S.-H. Lee et al., Nature (London) \textbf{418}, 856 (2002).
\bibitem{Radaelli02} P.G. Radaelli et al., Nature (London) \textbf{416}, 155 (2002).
\bibitem{Schmidt04} M. Schmidt et al., Phys. Rev. Lett. \textbf{92}, 056402 (2004).
\bibitem{Kondo97} S. Kondo et al., Phys. Rev. Lett. \textbf{78}, 3729 (1997).
\bibitem{Krimmel99} A. Krimmel et al., Phys. Rev. Lett. \textbf{82}, 2919 (1999).
\bibitem{Fritsch04} V. Fritsch et al., Phys. Rev. Lett. \textbf{92}, 116401 (2004).
\bibitem{Fichtl05} R. Fichtl et al., Phys. Rev. Lett. \textbf{94}, 027601 (2005).
\bibitem{Tsurkan05} V. Tsurkan et al., J. Phys. Chem. Solids \textbf{66}, 2036 (2005).
\bibitem{Lee00} S.-H. Lee et al., Phys. Rev. Lett. \textbf{84}, 3718 (2000).
\bibitem{Sushkov05} A.V. Sushkov et al., Phys. Rev. Lett. \textbf{94}, 137202 (2005).
\bibitem{Hemberger06} J. Hemberger et al., Phys. Rev. Lett.\textbf{97}, 087204 (2006).
\bibitem{Hemberger07} J. Hemberger et al., Phys. Rev. Lett. \textbf{98}, 147203 (2007).
\bibitem{Ueda05} H. Ueda et al., Phys. Rev. Lett. \textbf{94}, 47202 (2005).
\bibitem{Matsuda07} M. Matsuda et al., Nature Physics \textbf{3}, 397 (2007).
\bibitem{Yamashita00} Y. Yamashita et al., Phys. Rev. Lett. \textbf{85}, 4960 (2000).
\bibitem{Tchernyshyov02} O. Tchernyshyov et al., Phys. Rev. Lett. \textbf{88}, 067203 (2002).
\bibitem{Rudolf07} T. Rudolf et al., New J. Phys. \textbf{9}, 76 (2007).
\bibitem{Hamedoun86a} M. Hamedoun et al., J. Phys. C \textbf{19}, 1783 (1986).
\bibitem{Hamedoun86b} M. Hamedoun et al., J. Phys. C \textbf{19}, 1801 (1986).
\bibitem{Yokaichiya09} F. Yokaichiya et al., Phys. Rev. B \textbf{79}, 064423 (2009).
\bibitem{Ji09} S. Ji et al., Phys. Rev. Lett. \textbf{103}, 037201 (2009).
\bibitem{Lüthi05} B. L\"{u}thi, \emph{Physical Acoustics in the Solid State}, Springer, Berlin, (2005).
\bibitem{Kino71} Y. Kino and B. L\"{u}thi, Solid State Comm. \textbf{9}, 805 (1971).
\bibitem{Wolf01} B. Wolf et al., Physica B \textbf{294-295}, 612 (2001).
\bibitem{Varshni70} Y.P. Varshni, Phys. Rev. B \textbf{2}, 3952 (1970).
\bibitem{Zherlitsyn10} S. Zherlitsyn et al., J. Low Temp. Phys. \textbf{159}, 134 (2010).
\bibitem{Kojima10} E. Kojima et al., J. Low Temp. Phys. \textbf{159}, 3 (2010).
\bibitem{Mitamura07} H. Mitamura et al., J. Phys. Soc. Jpn. \textbf{76}, 085001 (2007).
\bibitem{Bhatta11} S. Bhattacharjee et al., Phys. Rev. B \textbf{83}, 184421 (2011).

\end{thebibliography}
\end{document}